\documentclass[sigconf]{acmart}
\usepackage{listings}
\usepackage{graphicx}
\usepackage{subcaption}

\usepackage{xcolor}
\usepackage{framed}

\copyrightyear{2020}
\acmYear{2020}
\setcopyright{acmlicensed}\acmConference[ESEM '20]{ESEM '20: ACM / IEEE International Symposium on Empirical Software Engineering and Measurement (ESEM)}{October 8--9, 2020}{Bari, Italy}
\acmBooktitle{ESEM '20: ACM / IEEE International Symposium on Empirical Software Engineering and Measurement (ESEM) (ESEM '20), October 8--9, 2020, Bari, Italy}
\acmPrice{15.00}
\acmDOI{10.1145/3382494.3410680}
\acmISBN{978-1-4503-7580-1/20/10}

\begin{document}

\title{A large-scale comparative analysis of Coding Standard conformance in Open-Source Data Science projects}

\author{Andrew J. Simmons}
\orcid{0000-0001-8402-2853}
\affiliation{%
  \institution{Deakin University}
  \department{Applied Artificial Intelligence Inst.}
  \streetaddress{Locked Bag 20000}
  \city{Geelong}
  \state{VIC}
  \country{Australia}
  \postcode{3220}
}
\email{a.simmons@deakin.edu.au}

\author{Scott Barnett}
\affiliation{%
  \institution{Deakin University}
  \department{Applied Artificial Intelligence Inst.}
  \streetaddress{Locked Bag 20000}
  \city{Geelong}
  \state{VIC}
  \country{Australia}
  \postcode{3220}
}
\email{scott.barnett@deakin.edu.au}

\author{Jessica Rivera-Villicana}
\affiliation{%
  \institution{Deakin University}
  \department{Applied Artificial Intelligence Inst.}
  \streetaddress{Locked Bag 20000}
  \city{Geelong}
  \state{VIC}
  \country{Australia}
  \postcode{3220}
}
\email{jessica.riveravillicana@deakin.edu.au}

\author{Akshat Bajaj}
\affiliation{%
  \institution{Deakin University}
  \department{Applied Artificial Intelligence Inst.}
  \streetaddress{Locked Bag 20000}
  \city{Geelong}
  \state{VIC}
  \country{Australia}
  \postcode{3220}
}
\email{akshat.bajaj@deakin.edu.au}


\author{Rajesh Vasa}
\affiliation{%
  \institution{Deakin University}
  \department{Applied Artificial Intelligence Inst.}
  \streetaddress{Locked Bag 20000}
  \city{Geelong}
  \state{VIC}
  \country{Australia}
  \postcode{3220}
}
\email{rajesh.vasa@deakin.edu.au}

\renewcommand{\shortauthors}{Simmons et al.}

\begin{abstract}
\textit{Background:} Meeting the growing industry demand for Data Science requires cross-disciplinary teams that can translate machine learning research into production-ready code. Software engineering teams value adherence to coding standards as an indication of code readability, maintainability, and developer expertise. However, there are no large-scale empirical studies of coding standards focused specifically on Data Science projects. \textit{Aims:} This study investigates the extent to which Data Science projects follow code standards. In particular, \textit{which standards are followed, which are ignored, and how does this differ to traditional software projects?} \textit{Method:} We compare a corpus of 1048 Open-Source Data Science projects to a reference group of 1099 non-Data Science projects with a similar level of quality and maturity. \textit{Results:} Data Science projects suffer from a significantly higher rate of functions that use an excessive numbers of parameters and local variables. Data Science projects also follow different variable naming conventions to non-Data Science projects. \textit{Conclusions:} The differences indicate that Data Science codebases are distinct from traditional software codebases and do not follow traditional software engineering conventions. Our conjecture is that this may be because traditional software engineering conventions are inappropriate in the context of Data Science projects.

\end{abstract}




\maketitle

\graphicspath{{figures/}}

\section{Introduction}
\label{sec:introduction}

Software that conforms to coding standards creates a shared understanding between developers \cite{Lee2015}, and improves maintainability \cite{DosSantos2018, Nundhapana2018}. Consistent coding standards also enhances the readability of the code by defining where to declare instance variables; how to name classes, methods, and variables; how to structure the code and other similar guidelines \cite{martin2009clean}. 

A growing trend in the software industry is the inclusion of Data Scientists on software engineering teams to build predictive models, analyze product and customer data, present insights to business and to build data engineering pipelines \cite{Kim2018DataChallenges}. Data Scientists may not follow common coding standards as they often come from non-software engineering backgrounds where software engineering best practices are unknown. Furthermore, a Data Scientist focuses on producing insights from data and building predictive models rather than on the inherent quality attributes of the code. In this context of a mixed team of Data Scientists and Software Engineers, team members must collaborate to produce a coherent software artefact and a consistent coding standard is essential for cohesive collaboration and to maintain productivity. However, currently it is unknown if Data Scientists follow common coding standards. 

Recent studies have focused on conformance of coding standards applied to application categories \cite{zou2019does, Bafatakis2019, Papamichail2020fixed} and has not considered the inherent technical domain \cite{Barnett2017} of Data Science. For example, Data Science applications may follow implicit or undocumented guidelines for 1) describing mathematical notation when implementing algorithms , e.g. ``X'' is often used as the input matrix in statistics, 2) define domain specific constructs such as a ``models'' folder for storing predictive models, and 3) writing vectorised code rather than using an imperative style to improve performance. Thus, general coding standards may not be applicable to Data Science code. 

While research into improving the predictive accuracy of machine learning algorithms for Data Science has flourished, less attention has been paid to the the software engineering concerns of Data Science. Industry has reported growing architectural debt in machine learning, well beyond that experienced in traditional software projects \cite{Sculley2015}. While the primary concern is ``hidden'' system-level architectural debt, there may also be observable symptoms such as dead code paths due to rapid experimentation \cite{Sculley2015}. 

In this study we conduct an empirical analysis to investigate the discrepancies in conformance to coding standards between Data Science projects and traditional software engineering projects. The novel contributions of this research are threefold. First, to the best of the authors' knowledge this is the first large scale (1048 Open-Source Python Data Science projects and 1099 non-Data Science projects) empirical analysis of coding standards in Data Science projects. Second, we empirically identify differences in coding standard conformance between Data Science and non-Data Science projects. Finally, we investigate the impact of using machine learning frameworks on coding standards. 
Our study focuses on Python projects as this is the most popular language for Open-Source Data Science projects on GitHub \cite{Braiek2018}. 

Our work contributes empirical evidence towards understanding phenomena that data science teams experience anecdotally, and lays the foundation for future research to provide guidelines and tooling to support the unique nature of Data Science projects.

\section{Motivating Example}
\label{sec:motivation}

\begin{figure}[H]
  \includegraphics[width=0.9\linewidth]{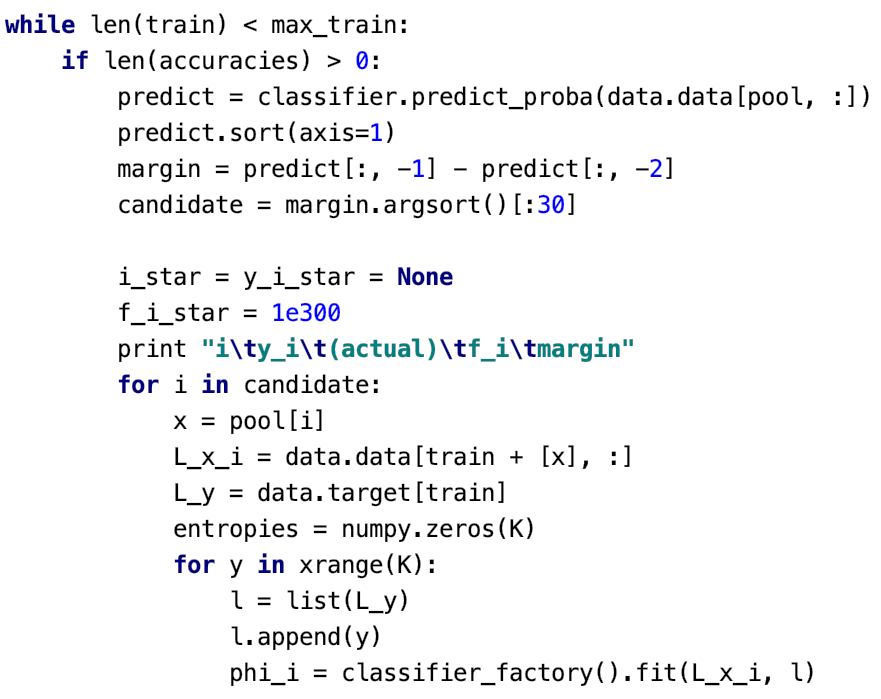}
  \caption[Code snippet]{Example code snippet from a machine learning loop taken from an Open-Source project\footnotemark}
  \label{fig:source_code}
\end{figure}

As a motivating example for our research, consider Naomi, a Data Scientist working with a team of other Data Scientists and Software Engineers to build a predictive model that will be deployed as part of the backend for a web application. She finds a recent research paper in the field and implements the latest algorithm using Python due to the availability of machine learning frameworks \cite{Braiek2018}. However, during automated code review Naomi finds out that her code does not adhere to the coding standards enforced by the web framework used within Naomi's company. A snippet of the kind of code written by Namoi is shown in Figure~\ref{fig:source_code}.

\footnotetext{https://github.com/shuyo/iir/blob/master/activelearn/mmms.py\#L18}

The team realises that there are exceptions to the coding standards they have put in place but would like to formally define these standards for consistency across the organisation. They also do not have any guidelines for implementing Data Science specific coding standards or where these standards should be applied. To assist Noami's team we investigate the answers to the following research questions (RQ):
\begin{itemize}

\item [\textbf{RQ1.}] What is the current landscape of Data Science projects? (Age, LOC, Cyclomatic Complexity, Distributions)
\item [\textbf{RQ2.}] How does adherence to coding standards differ between Data Science projects compared to traditional projects?
\item [\textbf{RQ3.}] Where do coding standard violations occur within Data Science projects?
\end{itemize}



\section{Background and Related Work}
\label{sec:conventions}
\subsection{Coding standards}

Coding standards, also known as code conventions and coding style, have been developed under the premise that consistent standards such as inline documentation, class and variable naming standards and organisation of structural constructs improve readability, assist in identifying incorrect code, and are more likely to follow best practices. These attributes are associated with improved maintainability \cite{spolsky2008more, posnett2011simpler}, readability \cite{DosSantos2018} and understandability \cite{Nundhapana2018} in general; however, certain practices may have negligible or even negative effect \cite{Lee2015}.


In this paper, we specifically discuss Python coding standards, given that this is the preferred language for Data Science projects \cite{Braiek2018}. The Style Guide for Python Code, known as PEP 8, includes standards regarding code layout, naming standards, programming recommendations and comments~\cite{Python.org2019}. In addition, standards for Python also cover package organisation and the use of virtual environments~\cite{allbee2018hands}. The Python community also strives to have a single way of doing things as defined in the `Zen of Python' \cite{Alexandru2018} so focus on analysing the official style guide for Python, PEP 8\footnote{https://www.python.org/dev/peps/pep-0008/}. Static analysis tools for Python programs is notoriously difficult due to the dynamic type system where types are determined at run-time rather than at compile time.  

\lstset{language=python, basicstyle=\ttfamily,keywordstyle={}, literate={\_}{}{0\discretionary{\_}{}{\_}}}

\label{sec:relatedwork}


\subsection{Mining of coding standards}

\citeauthor{Allamanis2014} designed a tool to infer code style based on examples in a domain and to recommend suggestions on how to improve the code consistency ~\cite{Allamanis2014}. \citeauthor{Markovtsev2019} later presented a fully-automated approach powered by a decision tree forest model~\cite{Markovtsev2019}.
~\citeauthor{zhu2014patterns} performed a study on the use of folders in software projects and its possible relationships with project popularity~\cite{zhu2014patterns}.
Using the premise that conventions will make code easier to read, understand and maintain, ~\citeauthor{smit2011code} defined a metric for ``convention adherence'' and analysed a set of projects to study consistency among teams regarding their code conventions~\cite{smit2011code}.
~\citeauthor{Jarczyk2014fixed} devised metrics for a project's popularity and the quality of user support by (virtual or remote) team members. These metrics were used to analyse possible correlations between project quality and the characteristics of the team that developed it~\cite{Jarczyk2014fixed}.
~\citeauthor{Raghuraman2019} performed a quantitative analysis on the relationship between UML model design and the quality of the project according to the number of issues in its repository. Their findings report that projects that were designed with UML have fewer software defects than projects designed without UML~\cite{Raghuraman2019}.
The ecosystem for analysis of static programming languages is mature; however, analysis of dynamic programming languages such as Python presents challenges. In our work, we analyse the Python programming language as this has become the \textit{lingua franca} for Data Science teams.

\subsection{Mining of Python repositories}
~\citeauthor{Bafatakis2019} used Pylint to analyse code-style conformance of Python snippets shared on the Stack Overflow question-answer website, and studied the relationship between code-style and vote score~\cite{Bafatakis2019}.
\citeauthor{Omari2019fixed} curated a corpus of 132 open source Python projects and analysed their Pylint scores and cyclomatic complexity~\cite{Omari2019fixed}. Their corpus includes machine learning repositories; however, it is dominated by web frameworks and all topics were analysed together as part of the same group.
\citeauthor{Biswas2019} formed a corpus of Data Science projects written in Python and made the dataset available using the Boa infrastructure \cite{Biswas2019}; however, as the Boa infrastructure only holds the abstract syntax tree rather than the underlying source code, their dataset cannot be used to analyse presentation related style violations as is.

As \citeauthor{Biswas2019} claimed to provide ``the first dataset that includes Data Science projects written in Python''~\cite{Biswas2019}, but do not provide a means for measuring presentation related aspects of code style; to the best of our knowledge, we provide the first large-scale empirical investigation to specifically examine the extent to which Data Science projects written in Python follow code standards.

\section{Empirical Study Design}
\label{sec:methodology}

\begin{figure}
  \includegraphics[width=\linewidth]{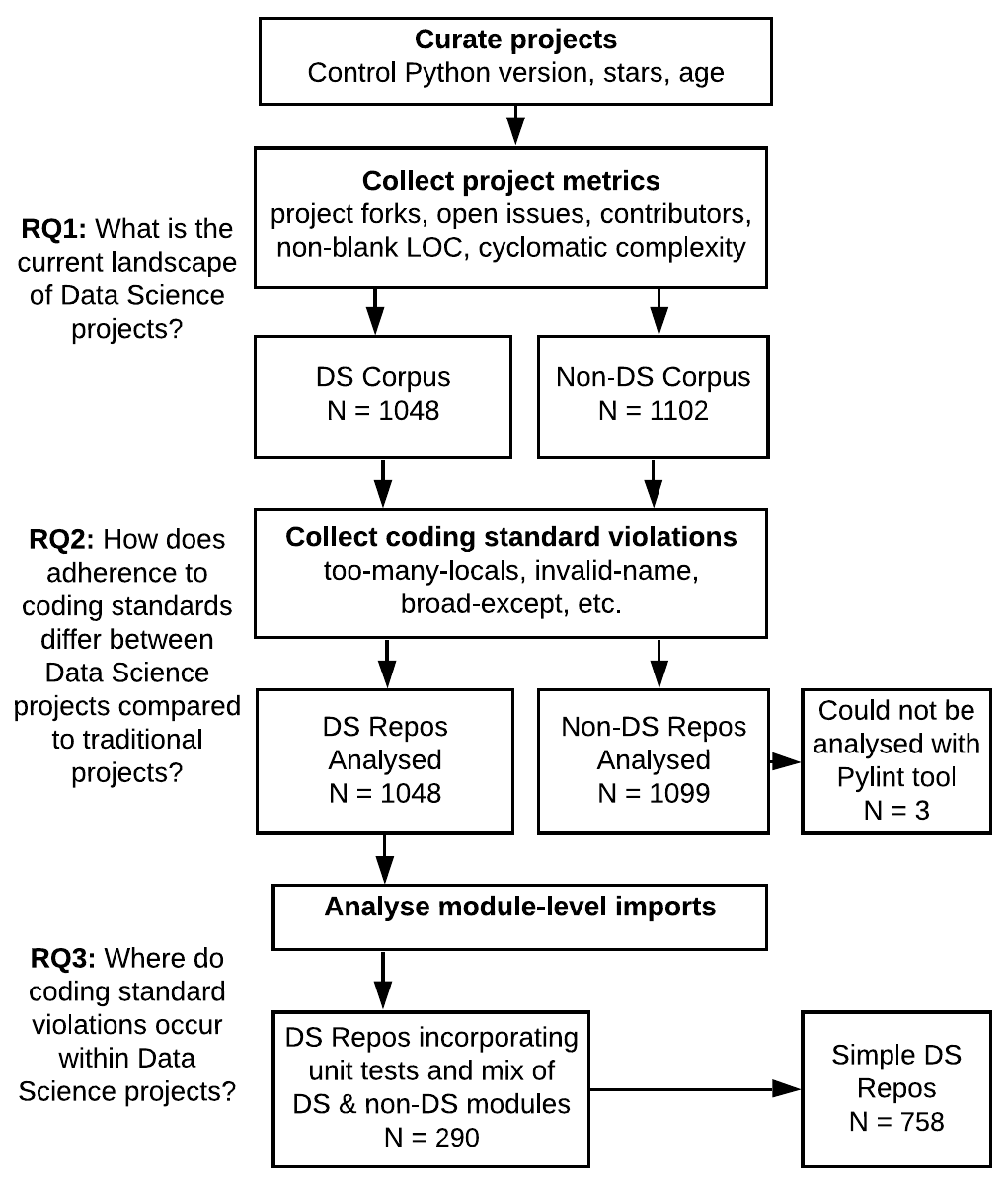}
  \caption{Study Design}
  \label{fig:methodology}
\end{figure}

To investigate the current landscape of data science projects (RQ1), we sought to use an established corpus of high-quality Data Science Projects. For this, we reuse the Boa Data Science corpus of 1558 Open-Source Python projects curated from GitHub by Biswas et al. \cite{Biswas2019} and recently presented at the \textit{Mining Software Repositories} (MSR2019) conference. However, we encountered two obstacles in attempting to reuse the existing corpus as-is. Firstly, Biswas et al. represent project code in the form of an abstract syntax tree for querying using the Boa platform. This discards the software syntax, which may hold valuable insights about developer behaviours \cite{Spinellis2011}. Secondly, understanding the unique aspects of Data Science projects requires a baseline of non-Data Science projects to contrast them against. As such, we curated a secondary corpus of non-Data Science projects, while controlling for project quality and project maturity in order to ensure a meaningful comparison of how adherence to coding standards differs between the two corpora (RQ2). A subset of the projects in the Data Science corpus were investigated at a deeper level to examine how coding standards vary between modules within projects (RQ3). The high level study design is presented in \autoref{fig:methodology}.

Results of the analysis, including raw logs of all detected code violations and source code to reproduce all results in this paper are made publicly available.\footnote{\url{https://doi.org/10.6084/m9.figshare.12377237.v2}}

\subsection{Formation of corpora}

\begin{figure}
  \includegraphics[width=\linewidth]{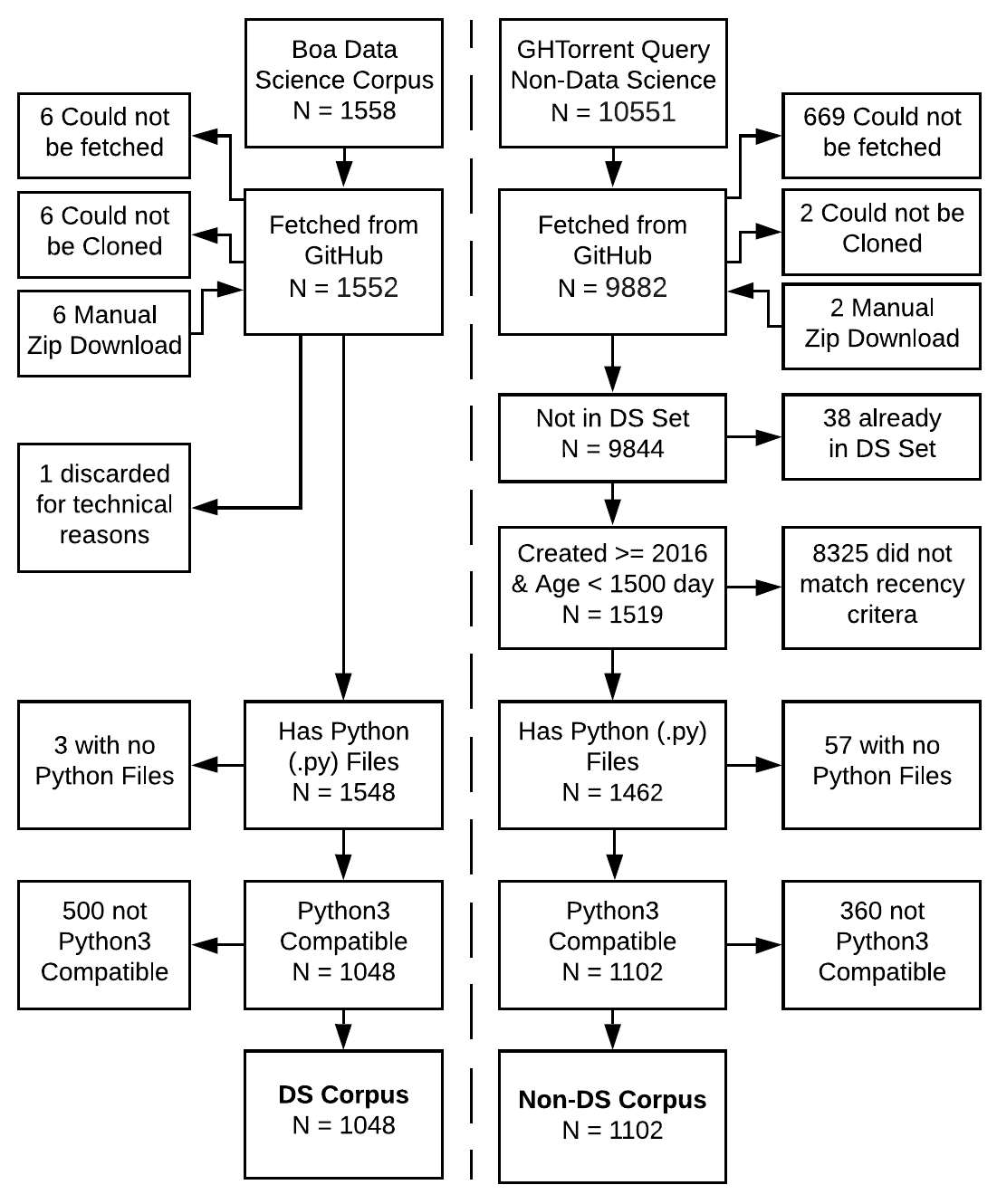}
  \caption{Curation of Projects}
  \label{fig:project_selection}
\end{figure}

\autoref{fig:project_selection} shows the number of project repositories available (1558 projects in the Boa Data Science corpus, and 10551 potential non-Data Science projects available on GitHub), and how we curated a selection of 1048 Data Science projects and 1102 Non-Data Science projects for further analysis.

\subsubsection{Data Science Corpus}
\label{sec:data-selection}

For each project in the Boa Data Science corpus, we fetched project metadata, such as the repository URL, current number of project stars, forks, issues, and committers, from the GitHub API. We were unable to fetch metadata from GitHub for six (0.4\%) of the projects listed in the Boa Corpus. This can occur due to projects on GitHub being renamed, deleted, or marked private by their owners after the publication of the Boa Data Science Corpus. We were able to retrieve full source code for all projects available via the GitHub API; however, we needed to manually download and extract six of the projects that could not be cloned using the Git client (as the downloaded zip archives did not include the project Git history, these six repositories were excluded from the age calculation and assumed to follow the same distribution as the other repositories in the corpus).

We needed to discard one repository in the corpus for technical reasons (we discovered that one of the files in the project was triggering a bug in Pylint that caused our analysis pipeline to hang indefinitely). Three of the projects were excluded as they contained no Python (.py) code files; while use of Python was a criteria checked by Biswas et al., it is possible that the Python files had since been deleted in the most recent version, or that the project had been replaced by another. Finally, we restricted the analysis to only Python 3 compatible projects. Python version 3 introduces major, backwards-incompatible changes to the Python language, which has impacted the entire Python ecosystem. Limiting the analysis to a single Python version helps to prevent confounding factors that would be introduced by mixing different Python versions. Furthermore, Python 2 is no longer supported, and recent versions of code analysis tools such as Pylint have dropped support for Python 2 code. To detect the Python version, we attempted to parse each Python (.py) file using both Python 2.7.13 and Python 3.6.5. If at least as many files can be parsed using Python 3 as can be parsed using Python 2, then we consider the project to be Python 3 compatible. It is also possible that repositories may include some Python files that cannot be parsed with either version of Python: this can occur due to syntax errors committed to the repository; use of syntax features only recently introduced in the latest versions of Python; or due to inclusion of special syntax intended for alternative environments such as Jupyter Notebooks. We included such repositories in the analysis, but only processed files in the repository that were compatible with Python 3 analysis tools (specifically, Pylint).



\subsubsection{Traditional (Non-Data Science) Corpus}
\label{sec:non-data-sci-selection}

To identify a corpus of traditional software repositories, we use the same project quality criteria that Biswas et al. \cite{Biswas2019} used to construct the Boa Data Science corpus, but negate any keywords relating to Data Science. As this corpus contains all types of projects other than Data Science, we will refer to it as the ``non-Data Science corpus''. Following Biswas et al. we select projects having at least 80 stars\footnote{GitHub allows users to star a project, thus serving as an indication to other users of the project's popularity and community acceptance. While stars may be subject to some manipulation, for the purpose of this paper we only use stars as a means to ensure that the Data Science and non-Data Science corpora are of a similar nature.} on GitHub (as an indication of community acceptance of the project quality), and that are not forked. In contrast to Biswas et al., we check that the project description does \textit{not} match any of their Data Science keywords, but still verify that the project has a non-empty description. Biswas et al. do an additional check that the project imports a common Data Science library. We considered negating this check; however, as their list includes some common libraries that could be used outside of a Data Science context (e.g. \texttt{matplotlib}, a general purpose data visualisation library), we decided to include projects regardless of their imports, so long as they weren't already included as part of Biswas et al.'s corpus.

Due to limitations imposed by the GitHub search API (search query results are limited to return at most 1000 repositories), we ran our search query against GHTorrent \cite{Gousios2013} (an archive of GitHub metadata). As Biswas et al. \cite{Biswas2019} do not state the time of their query, nor how they overcame the GitHub API limits, we chose to use the \textit{ght\_2018\_04\_01} GHTorrent dump available via Google BigQuery\footnote{\url{https://ghtorrent.org/gcloud.html}}. This query led to the identification of 10551 repositories. Of these, we were able to retrieve current metadata for 9882 repositories directly from the GitHub API, but unable to fetch 669 (6.8\%). Gousios notes  limitations in the way that GHTorrent collects data from GitHub \cite{Gousios2013} (e.g. that GHTorrent additively collects data from GitHub, but GitHub does not report deletions) that may explain why some repositories were listed in GHTorrent, but not available from GitHub. These effects would have been exasperated by repository changes since the time of the data dump. We were able to successfully obtain full source code for all the repositories for which the GitHub API returned metadata (albeit that two of these repositories needed to be manually extracted from zip archives). We removed a further 38 repositories that were already present in the Boa Data Science Corpus; while in theory our query should have excluded these, it may have been due to changes in Data Science keywords in the project description between the time of the GHTorrent dump and the time that Biswas et al. constructed the Boa Data Science Corpus.

As Data Science has recently undergone rapid growth \cite{Braiek2018}, it is also necessary to consider the maturity of projects in our non-Data Science set in order to ensure a meaningful comparison with Data Science projects. To achieve this, we only included projects that according to the GitHub API had been created in 2016 or later. However, we found cases of mature projects that matched this criterion due to being recently migrated to GitHub. Thus as an additional measure we define a project's age as the number of days between the first commit in the Git log and the most recent commit to the default branch, and restrict the selection to projects under 1500 days old (approx. 4.1 years). This reduced the non-Data Science set to 1519 repositories. After removing projects that contained no Python files, or were not Python 3 compatible, we were left with 1102 non-Data Science repositories (as discussed in \autoref{sec:collecting-convention-violations}, a further three were discarded in the coding standard analysis stage as they didn't contain any files that could be analysed).

\subsection{Overview of Corpora}
\label{sec:overview-of-corpora}

\begin{figure}[htpb]
  \centering
  \includegraphics[width=0.8\linewidth]{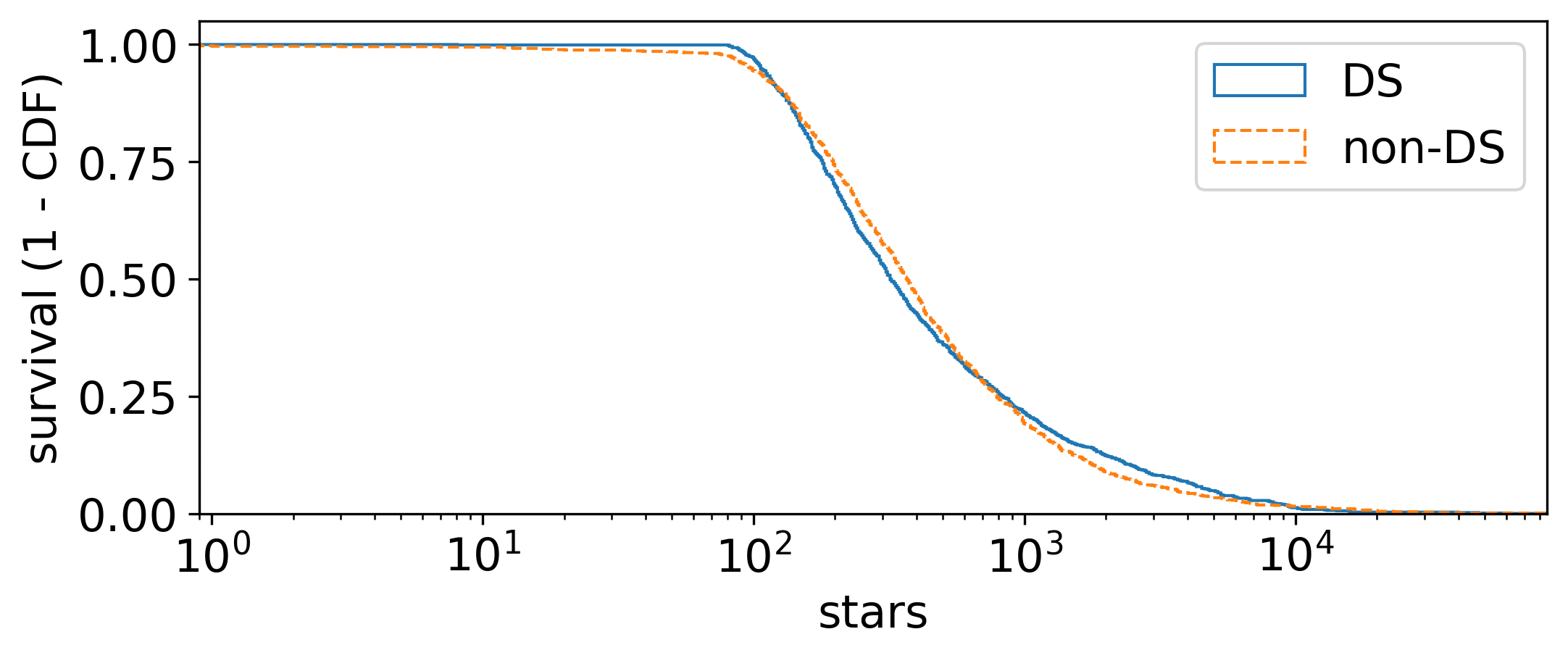}
  \includegraphics[width=0.8\linewidth]{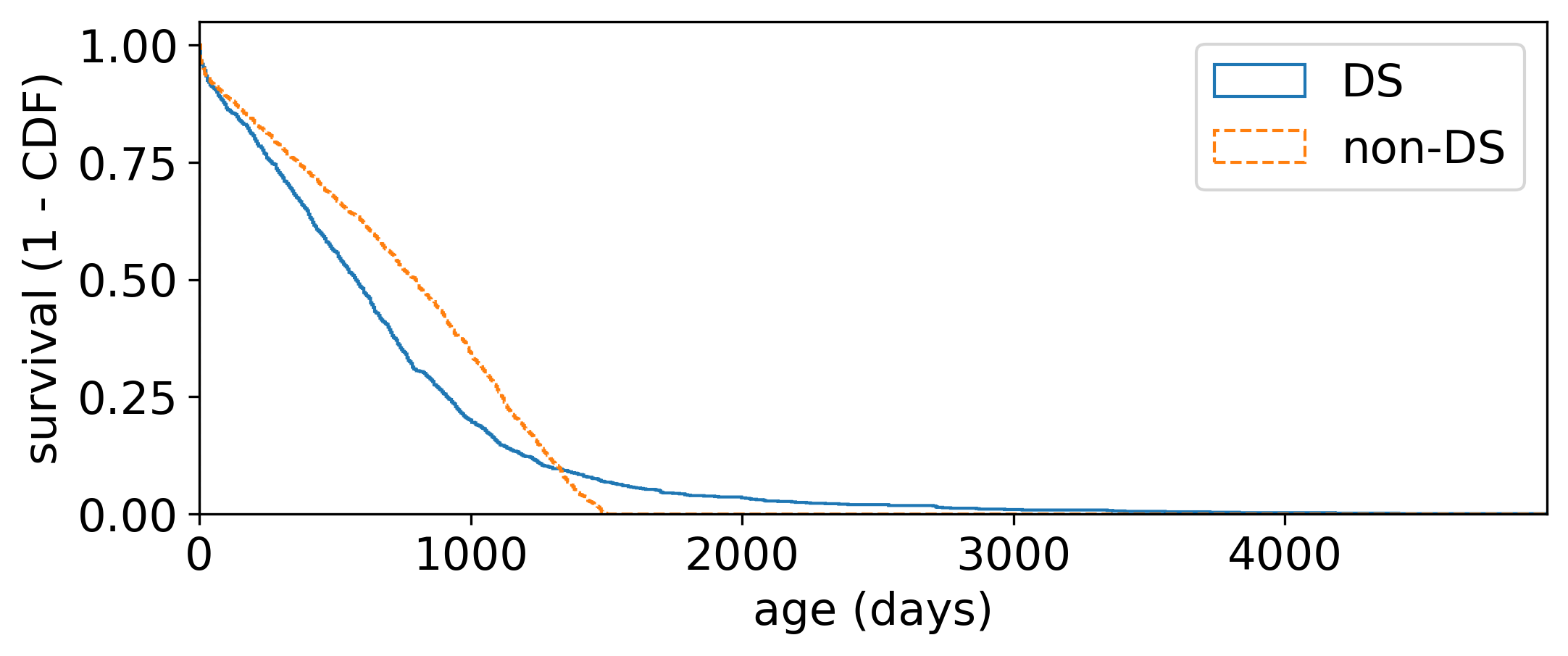}
  \caption{Survival plots showing controlled characteristics (stars, age) of the projects in the DS corpus and the Non-DS corpus (after selection). Stars were controlled for almost perfectly; however, age differences were only partially controlled for.}
  \label{fig:controlled-distributions}
\end{figure}

\autoref{fig:controlled-distributions} shows survival plots (1 - \textit{cumulative distribution function}) of the distribution of project stars and project age (after selection). To quantify the differences between the two corpora, we use the two-sample Kolmogorov--Smirnov (K--S) statistic (computed using SciPy version 1.4.1\footnote{\url{https://docs.scipy.org/doc/scipy/reference/generated/scipy.stats.ks_2samp.html}}), which is defined as the maximum vertical difference between the two curves in the survival plots, and can be used to test the null hypothesis that the samples from the two corpora are drawn from the same statistical distribution.

The distribution of stars in both corpora is almost identical; the two-sample K--S statistic is 0.06, which represents only a weakly significant (p=0.03) difference beyond what would be expected by chance if drawn from the same distribution. In contrast, despite our attempt to restrict the age of non-Data Science projects to match that of the Data Science corpora, it is only an approximate fit (K--S statistic of 0.19). The mean age of the selected projects in the Data Science Corpus was 671 days, whereas after selection, the mean age of projects in the Non-Data Science corpus was 743 days. We could have imposed additional selection criteria on the Non-Data Science corpus to bring the age distribution closer; however, this would have reduced the size of the dataset and further shifted the distribution away from the underlying population of non-Data Science projects. Our intent was to ensure both corpora have similar project quality and maturity characteristics to permit a meaningful comparison of coding standards between the corpora rather than to perfectly control all characteristics.

\subsection{Collecting project metrics}
\label{sec:collecting-project-metrics}

In addition to project stars and age (defined in \autoref{sec:data-selection}), we examined how these relate to a broad range of project-level metrics such as forks, issues, contributors, lines of code, and cyclomatic complexity in order to understand the current landscape of Data Science projects. The number of \textbf{project forks}, \textbf{open issues}, and \textbf{contributors} were retrieved through the GitHub API along with the number of stars when constructing the corpus. After cloning the repositories, we calculated the \textbf{non-blank Lines Of Code (LOC)} for all Python files (anything ending in the \texttt{.py} extension). The advantage of the non-blank lines of code metric is that the metric is easy to define (all lines of code, including comments, other than lines that consist entirely of whitespace) and reproduce \cite{Malloy2019}. It is also fast to calculate (even for thousands of repositories), and doesn't require parsing the source code. We used Radon \cite{Radon2018} to compute \textbf{McCabe's cyclomatic complexity}; \cite{McCabe1976a} the number of linearly independent paths through the code, equivalent to the number of decisions in a code block plus 1~\cite{McCabe1976a}.


\subsection{Collecting coding standard violations}
\label{sec:collecting-convention-violations}

\textbf{Pylint} \cite{Pylint} is a tool to perform a series of error checks, attempt to enforce coding standards (as per PEP 8, discussed previously), and find code smells (i.e., suspicious code that may be an indication of deeper problems in the system~\cite{fowler2018refactoring}). Pylint analyses code and displays a set of messages, categorised as convention violations, refactors for code smells, errors or bugs, warnings for Python specific problems, and fatal errors that prevented Pylint from further processing. We executed Pylint on each module (file) of each repository in order to count the number of code standard violations of each type. Any files that could not be analysed (e.g. due to using or importing invalid syntax, broken symbolic links, or triggering internal Pylint issues) were removed from this stage of the analysis (including removal from the LOC counts to ensure they did not distort the frequency of warnings per LOC). This resulted in three non-Data Science repositories being discarded completely, leaving a total of 1048 Data Science repositories and 1099 non-Data Science repositories.

While many of the coding standards enforced by Pylint are configurable, we left these as their default values to reflect the community standards. Pylint gives the ability to ignore unwanted warning types, but for our purposes we included all warning types in the analysis then grouped them at the most detailed level. This allowed us to identify precisely which warning types are significant.

\subsection{Analysing module-level import graph}
\label{sec:analysing-module-level-import-graph}

We used the Python \textit{FindImports} library\footnote{\url{https://pypi.org/project/findimports/}} to extract the direct dependencies (imports) of each module (file). To detect unit tests, imports were checked against a list of common Python testing frameworks (\textit{unittest}, \textit{pytest}, \textit{unittest2}, \textit{mock}). We checked the remaining modules against a the same list of Data Science libraries as Biswas et al. \cite{Biswas2019} (machine learning, visualisation, and data wrangling libraries).

\section{Results}
\label{sec:analysis}

\subsection{(RQ1) What is the current landscape of Data Science projects?}

\begin{figure}[htpb]
  \centering
  \includegraphics[width=0.8\linewidth]{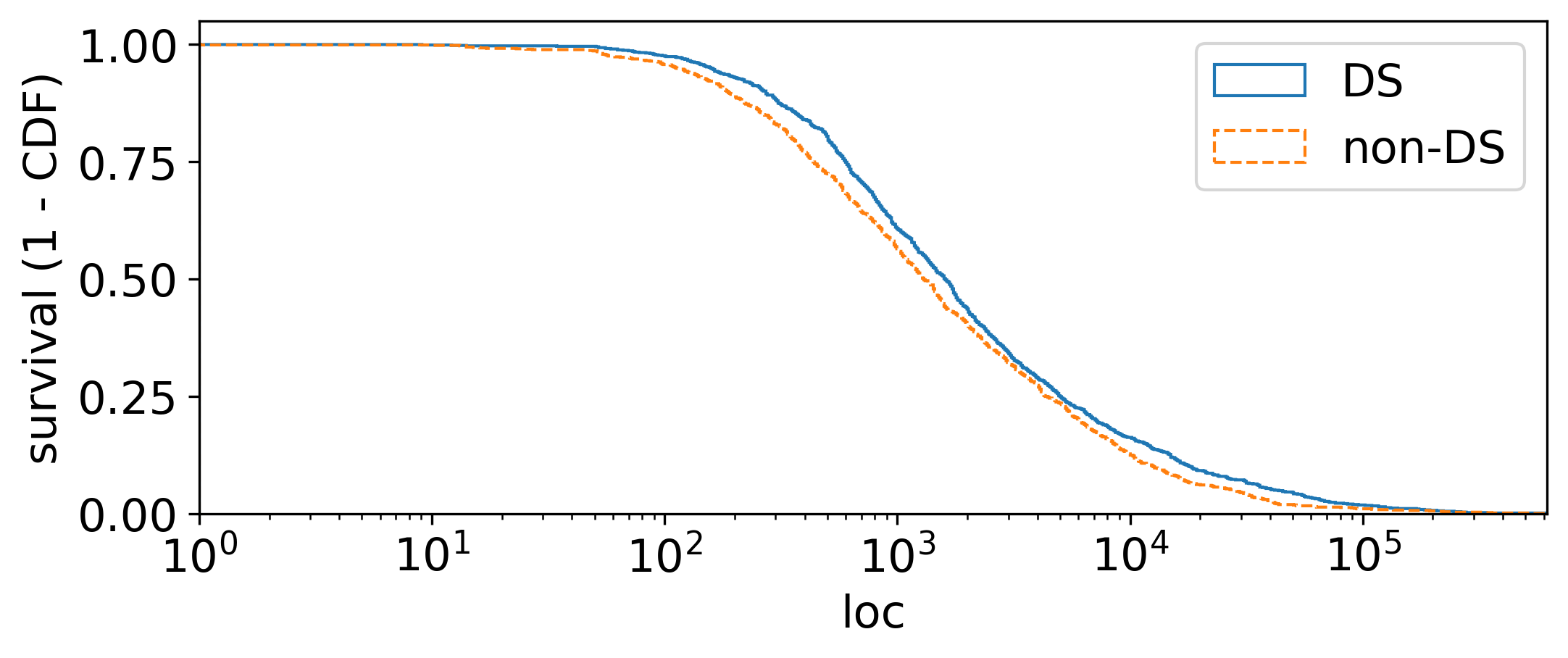}
  \includegraphics[width=0.8\linewidth]{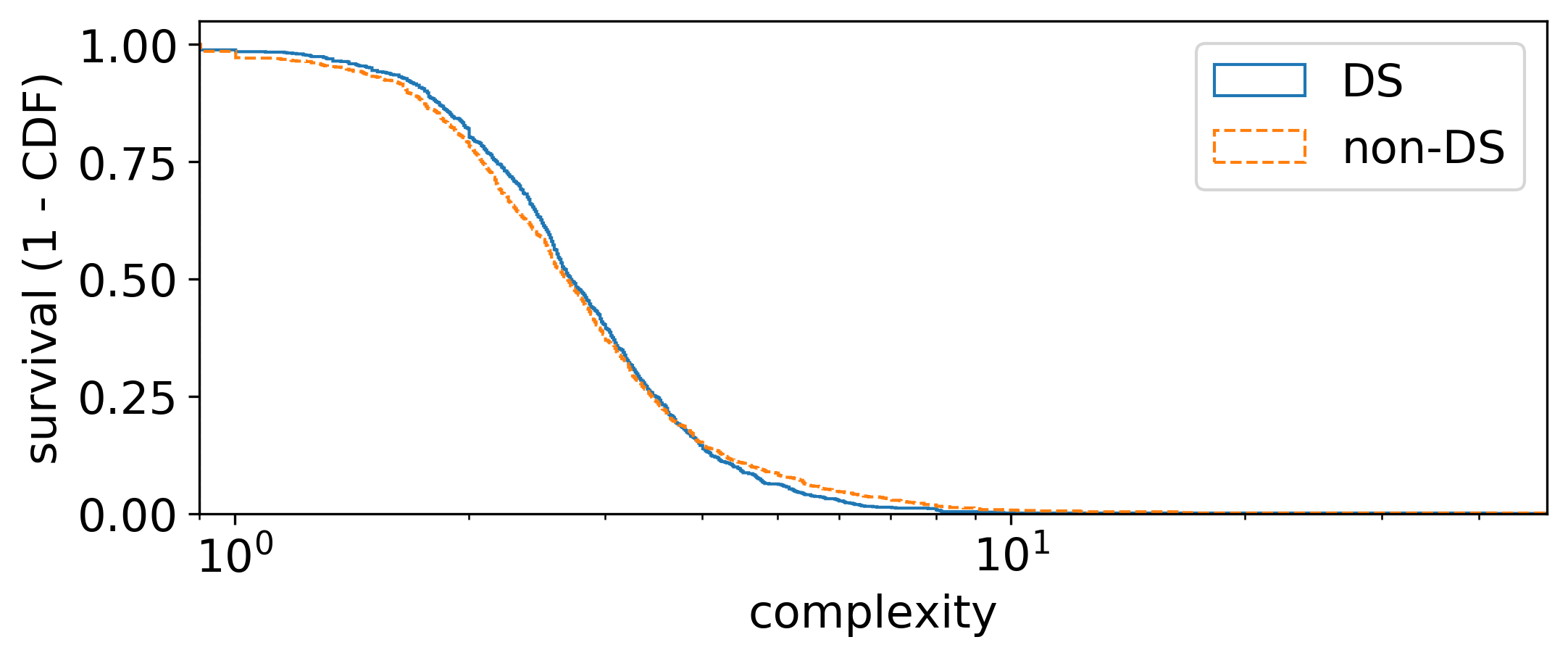}
  \includegraphics[width=0.8\linewidth]{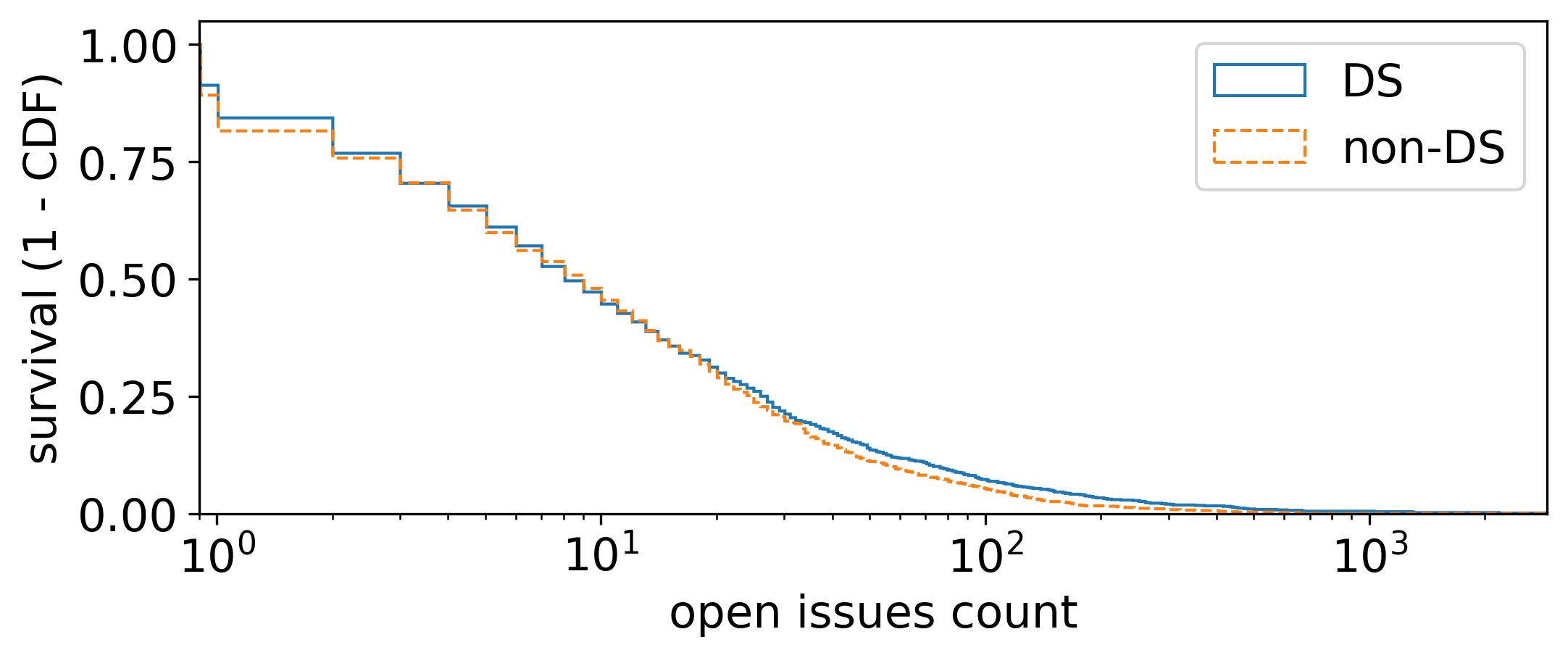}
  \includegraphics[width=0.8\linewidth]{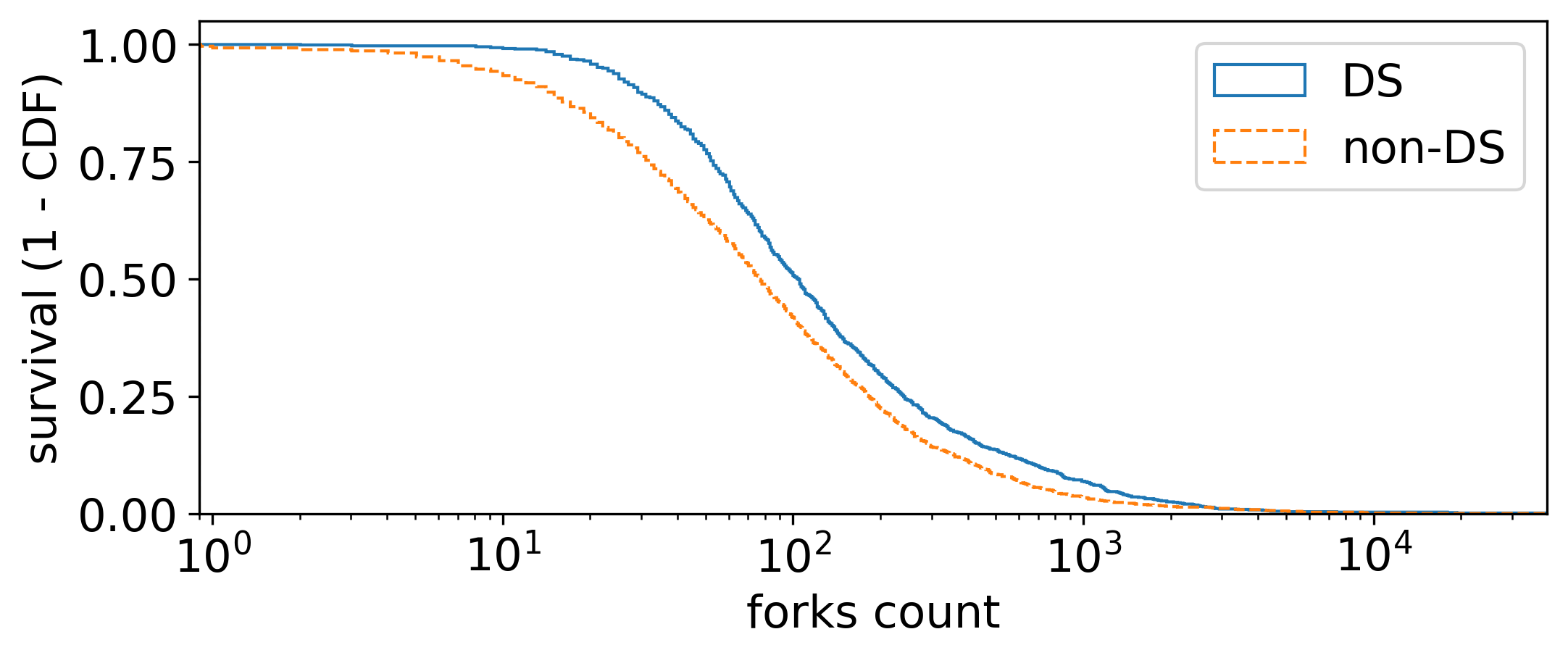}
  \includegraphics[width=0.8\linewidth]{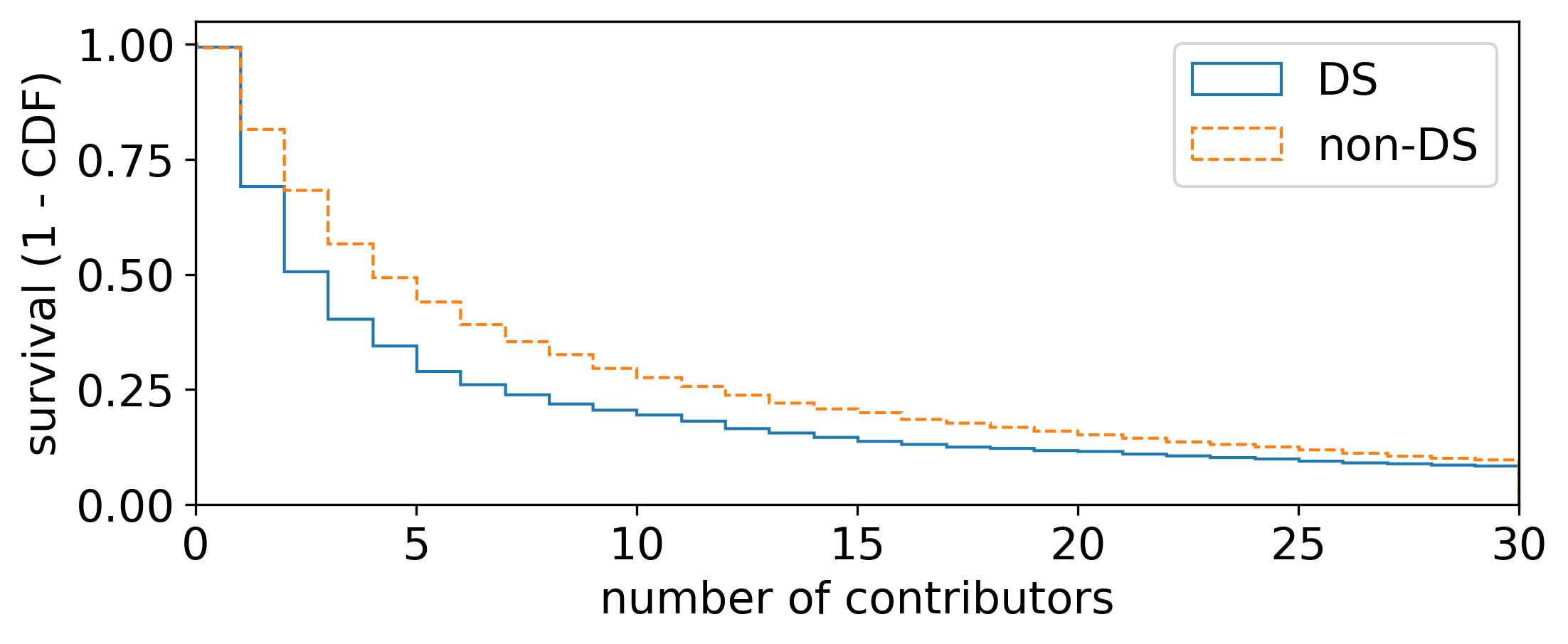}
  \caption{Survival plots showing characteristics of the projects selected for the DS corpus and the control group.}
  \label{fig:distributions}
\end{figure}

In this section, we investigate the Data Science landscape through the lens of the project level metrics described in \autoref{sec:collecting-project-metrics}, while controlling for the project stars (as an indicator of community acceptance) and age (as an indicator of project maturity) as described in \autoref{sec:overview-of-corpora} to ensure a meaningful comparison between the Data Science corpus and the non-Data Science corpus. The resultant distributions are presented in \autoref{fig:distributions}.

Lines of code was greater in the Data Science corpus. When manually examining the results, we found cases of projects that encoded data files as Python code in order to make data easy to access from Python, which may explain some of the difference, particularly in the tail of the distribution.


For each project, cyclomatic complexity was calculated for each method or function, then averaged to provide a score for that project (projects without any methods or functions were excluded from the analysis). Cyclomatic complexity is similar in both corpora (K--S statistic=0.06, p=0.05). This was an unexpected result, as implementations of machine learning algorithms are an important component of Data Science projects, which one would intuitively expect to result in a higher cyclomatic complexity caused by nested loops and branches. A potential explanation is the use of frameworks and vectorised code that will tend to flatten out loops.

The number of open issues shows no significant differences (K--S statistic=0.03, p=0.6) between Data Science and non-Data Science projects. While our selection process attempts to control the distribution of stars and age, we did not make any attempt to manipulate the distribution of open issues. The finding that open issues were also similar between the projects is an indication that the Data Science and non-Data Science corpora are of a similar quality and maturity.

Data Science projects had a lower number of contributors than non-Data Science projects, but had a higher number of forks. This is consistent with a development workflow whereby data scientists copy/fork each other's code/models, then work alone or in small teams to tune it to their use case. However further investigation would be needed to confirm whether this is the case.

\begin{framed}
\textbf{RQ1 summary.} Compared to other projects of similar quality (stars) and maturity (age), Data Science projects on GitHub have the same number of open issues and less contributors, but are forked more often and, surprisingly, have similar average cyclomatic complexity to non-Data Science projects.
\end{framed}

\subsection{(RQ2) How does adherence to coding standards differ between Data Science projects compared to traditional projects?}
\label{sec:pylint-ml}

\begin{table*}
\caption{Coding standard violations (warnings / non-blank LOC) found to significantly differ between DS repositories and Non-DS repositories. Largest values of mean and median highlighted in bold}
\label{tab:pylint-differences}
\begin{tabular}{lllllll}
\toprule
Pylint warning (per non-blank LOC) & DS mean  & DS median & Non-DS mean & Non-DS median & p-value \\
\midrule
too-many-locals                   & \textbf{0.34\%} & \textbf{0.28\%} & 0.16\% & 0.06\% & $1.10 \cdot 10^{-59}$ \\
import-error                      & \textbf{2.73\%} & \textbf{2.07\%} & 1.83\% & 1.16\% & $2.02 \cdot 10^{-45}$ \\
too-many-arguments                & \textbf{0.33\%} & \textbf{0.26\%} & 0.19\% & 0.08\% & $3.12 \cdot 10^{-42}$ \\
invalid-name                      & \textbf{9.88\%} & \textbf{7.73\%} & 6.94\% & 4.86\% & $1.31 \cdot 10^{-26}$ \\
broad-except                      &         0.03\%  &         0.00\%  & \textbf{0.13\%} & 0.00\% & $8.94 \cdot 10^{-20}$ \\
bad-indentation                   & \textbf{6.58\%} &         0.00\%  & 2.24\% & 0.00\% & $1.23 \cdot 10^{-14}$ \\
consider-using-enumerate          & \textbf{0.05\%} &         0.00\%  & 0.03\% & 0.00\% & $7.06 \cdot 10^{-14}$ \\
missing-class-docstring           &         0.43\%  &         0.26\%  & \textbf{0.73\%} & \textbf{0.44\%} & $6.17 \cdot 10^{-13}$ \\
trailing-whitespace               & \textbf{1.59\%} &         0.05\%  & 1.05\% & 0.00\% & $1.43 \cdot 10^{-12}$ \\
missing-function-docstring        &         2.56\%  &         2.31\%  & \textbf{3.26\%} & \textbf{2.94\%} & $6.02 \cdot 10^{-12}$ \\
unsubscriptable-object            & \textbf{0.04\%} &         0.00\%  & 0.02\% & 0.00\% & $7.44 \cdot 10^{-12}$ \\
inconsistent-return-statements    &         0.03\%  &         0.00\%  & \textbf{0.07\%} & 0.00\% & $1.35 \cdot 10^{-10}$ \\
wrong-import-position             & \textbf{0.32\%} &         0.00\%  & 0.18\% & 0.00\% & $2.54 \cdot 10^{-9}$ \\
bad-whitespace                    & \textbf{5.43\%} & \textbf{0.53\%} & 4.90\% & 0.15\% & $2.68 \cdot 10^{-9}$ \\
too-many-instance-attributes      & \textbf{0.11\%} & \textbf{0.05\%} & 0.08\% & 0.02\% & $4.82 \cdot 10^{-9}$ \\
unnecessary-comprehension         & \textbf{0.03\%} &         0.00\%  & 0.02\% & 0.00\% & $7.74 \cdot 10^{-9}$ \\
missing-final-newline             & \textbf{0.12\%} &         0.00\%  & 0.20\% & 0.00\% & $1.11 \cdot 10^{-8}$ \\
unused-variable                   & \textbf{0.38\%} & \textbf{0.23\%} & 0.33\% & 0.14\% & $8.46 \cdot 10^{-8}$ \\
ungrouped-imports                 & \textbf{0.05\%} &         0.00\%  & 0.04\% & 0.00\% & $1.81 \cdot 10^{-7}$ \\
too-few-public-methods            &         0.17\%  &         0.08\%  & \textbf{0.36\%} & \textbf{0.12\%} & $4.04 \cdot 10^{-7}$ \\
too-many-public-methods           &         0.00\%  &         0.00\%  & \textbf{0.01\%} & 0.00\% & $6.24 \cdot 10^{-7}$ \\
missing-module-docstring          &         0.82\%  &         0.65\%  & \textbf{1.17\%} & \textbf{0.74\%} & $7.03 \cdot 10^{-7}$ \\
too-many-statements               & \textbf{0.07\%} & \textbf{0.03\%} & 0.06\% & 0.00\% & $1.78 \cdot 10^{-6}$ \\
attribute-defined-outside-init    & \textbf{0.42\%} & \textbf{0.01\%} & 0.24\% & 0.00\% & $2.05 \cdot 10^{-6}$ \\
line-too-long                     & \textbf{2.54\%} & \textbf{1.64\%} & 2.08\% & 1.02\% & $3.45 \cdot 10^{-6}$ \\
bad-continuation                  &         2.17\%  & \textbf{0.51\%} & \textbf{2.22\%} & 0.34\% & $5.88 \cdot 10^{-6}$ \\
bad-option-value                  & \textbf{0.004\%} &        0.00\%  & 0.002\% & 0.00\% & $8.38 \cdot 10^{-6}$ \\
consider-using-dict-comprehension & \textbf{0.007\%} &        0.00\%  & 0.003\% & 0.00\% & $1.32 \cdot 10^{-5}$ \\
trailing-newlines                 & \textbf{0.09\%} &         0.00\%  & 0.07\% & 0.00\% & $4.00 \cdot 10^{-5}$ \\
invalid-unary-operand-type        & \textbf{0.003\%} &        0.00\%  & 0.001\% & 0.00\% & $8.94 \cdot 10^{-5}$ \\
reimported                        & \textbf{0.024\%} &        0.00\%  & 0.018\% & 0.00\% & $1.37 \cdot 10^{-4}$ \\
\bottomrule
\end{tabular}
\end{table*}

We analysed all files in each repository with Pylint (version 2.4.4) and computed the number of coding standard violations per non-blank LOC. Pylint checked source code for 229 different types of code standard violations. We then examined differences in the error distribution between Data Science repositories compared to non-Data Science repositories.

Due to the sparse nature of code violations, many (or in some cases most) repositories contained zero violations of a particular type, whereas those that contained violations tended to do so with an extreme frequency (e.g. projects that used non-standard formatting could trigger multiple violations per line). As such, a two-sided Mann-Whitney U-test (computed using SciPy version 1.4.1\footnote{\url{https://docs.scipy.org/doc/scipy/reference/generated/scipy.stats.mannwhitneyu.html}}) was selected to detect differences in the distributions, as it is a non-parametric test that depends only on the ranking (ordering) of values rather than their absolute quantities. To guard against data dredging (an inflated number of false positives as a consequence of performing many comparisons), we apply Bonferroni correction as a conservative counter-measure. We set a significance threshold of $p < 0.05 / 229  \approx 2.2\cdot 10^{-4}$.

\begin{figure}[htpb]
  \centering
  \includegraphics[width=0.8\linewidth]{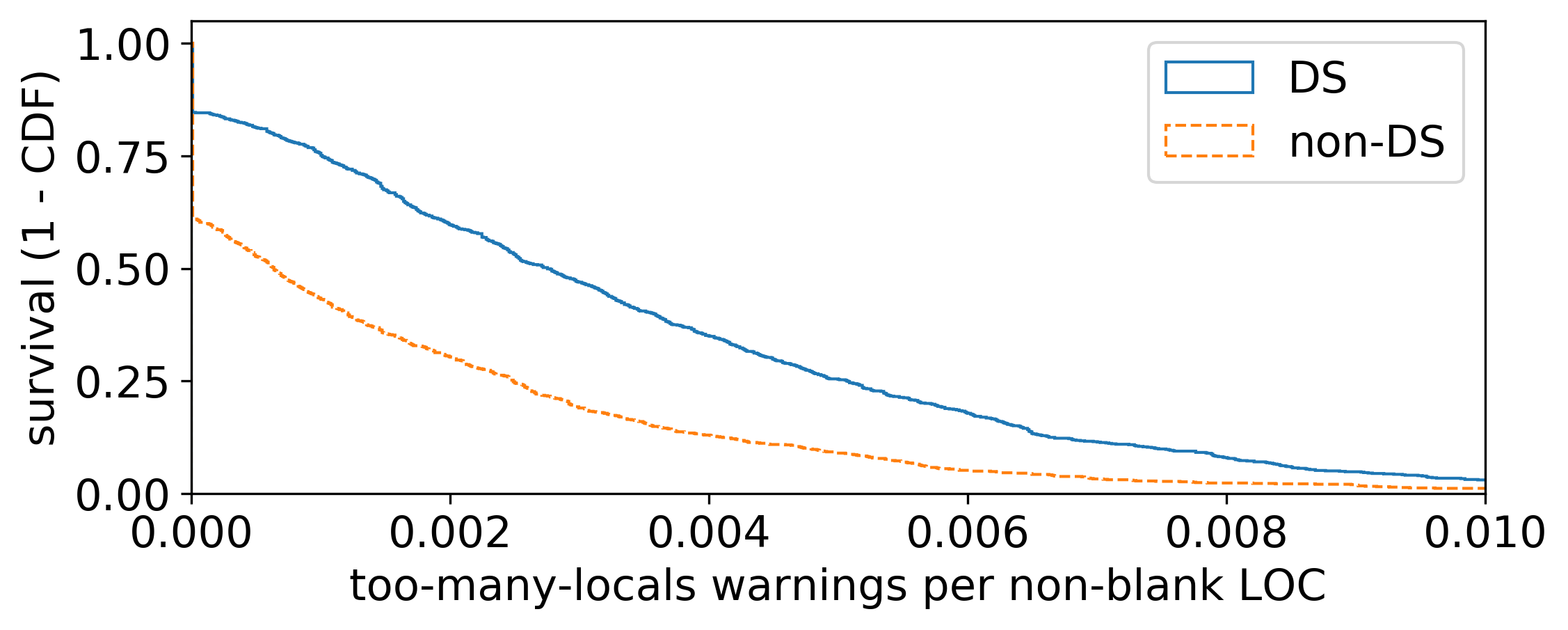}
  \includegraphics[width=0.8\linewidth]{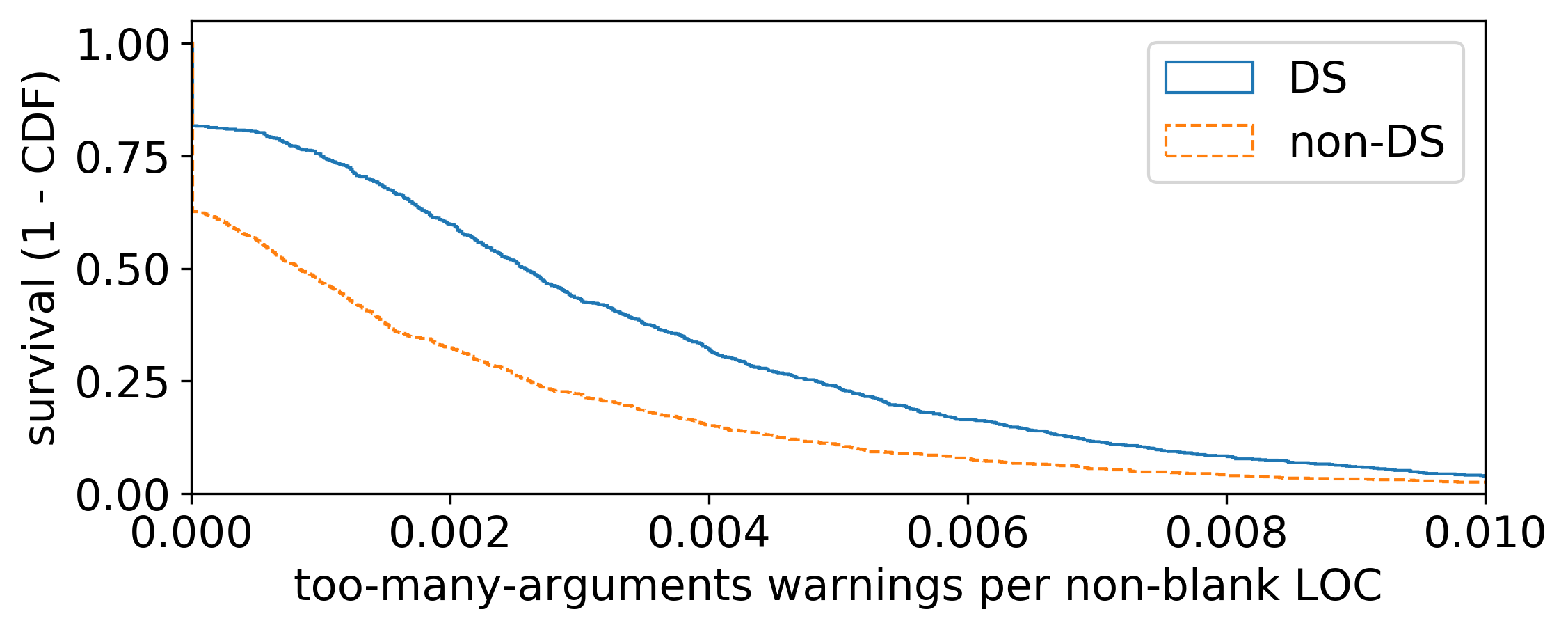}
  \includegraphics[width=0.8\linewidth]{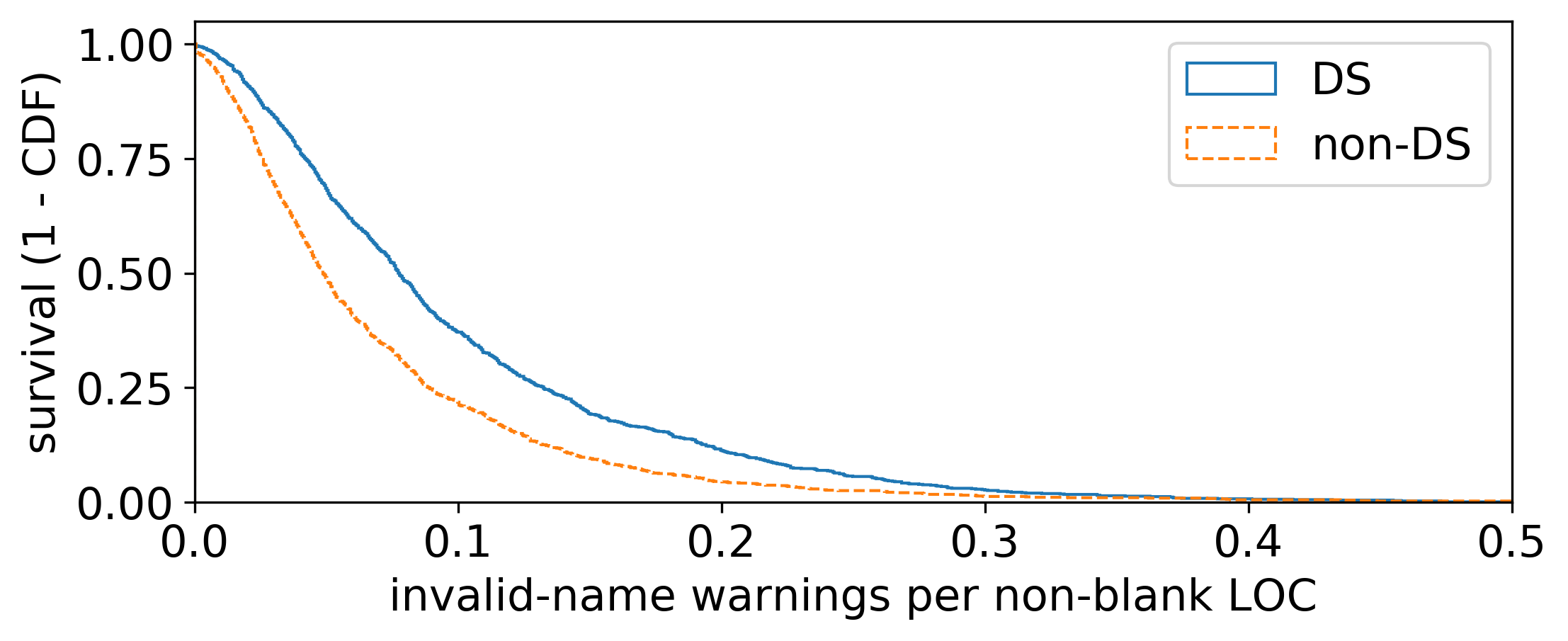}
  \includegraphics[width=0.8\linewidth]{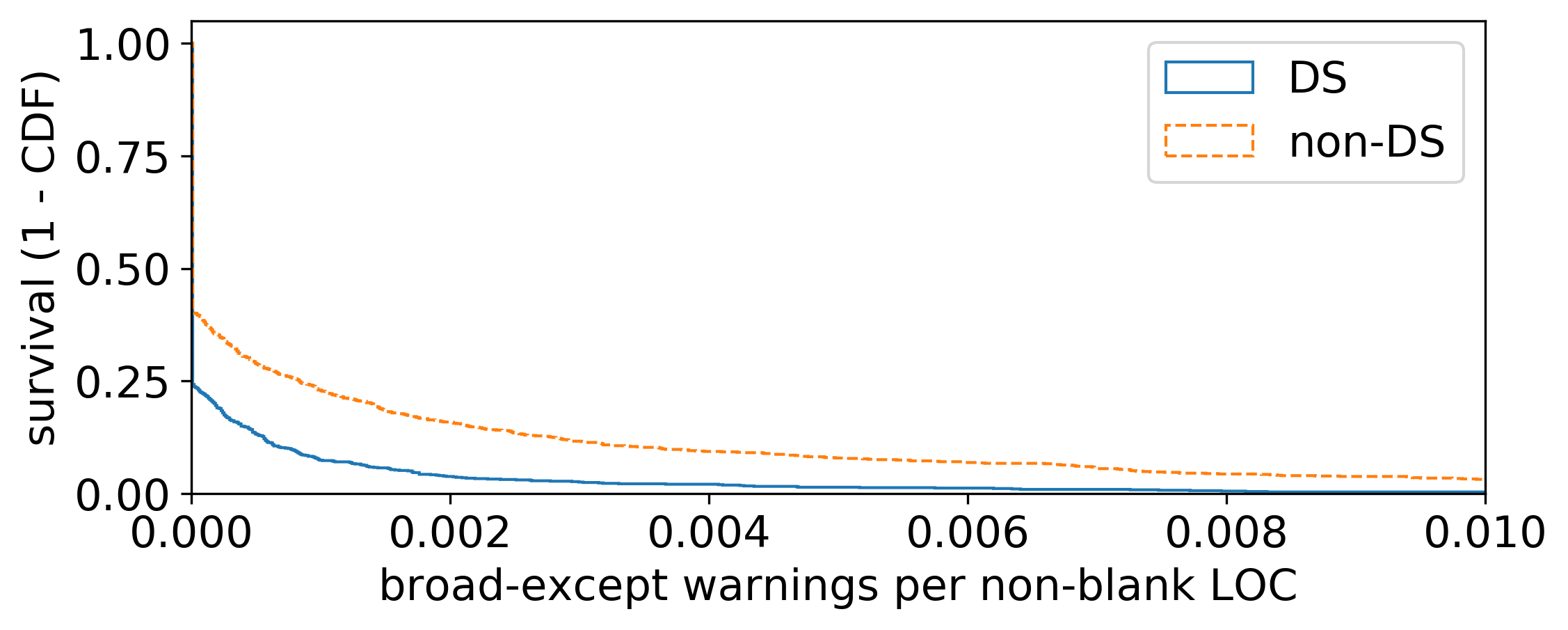}
  \caption{Survival plots comparing Data Science coding standard violation frequency to non-Data Science projects.}
  \label{fig:freq-pylint}
\end{figure}

A table of the Pylint warning types that had significantly different frequencies between Data Science applications compared to other projects is provided in \autoref{tab:pylint-differences}. The table shows that the top five most significant differences were: too-many-locals, import-error, too-many-arguments, invalid-name, and broad-except. The distributions of key warnings are compared visually in \autoref{fig:freq-pylint}.

Data Science projects (mean of 0.34\% warnings/LOC) trigger the \textit{too-many-locals} warning an order of magnitude more frequently than non-Data Science projects  (mean of 0.16\% warnings/LOC). By default, Pylint triggers the \textit{too-many-locals} warning when a function of method contains more than 15 variables. As function arguments count towards this limit, it is related to the \textit{too-many-arguments} warning which triggers when a function definition contains more than five parameters. Combined with our findings from RQ1, this implies that Data Science projects tend to incorporate functions with many variables; however, not necessarily a high cyclomatic complexity. A possible cause is models with multiple hyperparameters that are either hard-coded as variables in the function definition or passed to the the function as individual parameters rather than being stored in a configuration object.

Data Science projects also contained more frequent violations of Pylint's variable naming rules. Pylint will flag variables that are improperly capitalised (e.g. local variables are expected to be lowercase) or were too short (e.g. single letter variable names other than loop variables). This may conflict with mathematical notations, e.g. upper case "X" for an input matrix.

The \textit{import-error} difference and \textit{bad-option-value} warnings are artifacts due to Pylint that should be ignored. As we did not attempt to install individual project dependencies into the environment, the larger number of \textit{import-error}s triggered by Data Science projects are more likely an indication that they are importing external libraries more frequently.

\textit{broad-except} warning represents cases where a try-catch statement is used to catch a general (overly broad) exception rather than a particular type. This warning was more common in non-Data Science projects. However, the difference could also be explained by Data Science projects that fail to use exceptions handling at all. Similarly, the \textit{missing-class-docstring}, \textit{missing-function-docstring}, and \textit{missing-module-docstring} warnings were more common in non-Data Science projects; however, it is unclear whether this is due to worse documentation or a consequence of an Object-Oriented design with shorter modules and more common use of classes and functions.

\begin{framed}
\textbf{RQ2 summary.} Data Science projects contain over twice as many cases (per line of code) of functions/methods with excessive numbers of local variables. Data Science projects contain more frequent violations (per line of code) of variable naming standards than non-Data Science projects.
\end{framed}

\subsection{(RQ3) Where do coding standard violations occur within Data Science projects?}
\label{sec:module-level-analysis}

In this section, we examine how coding standard violations vary within projects. We first label any modules that import unit testing libraries as \textit{Unit Test}. The remaining modules are checked to see if they import any Data Science or Machine Learning libraries (e.g. \textit{tensorflow}) and labelled as either \textit{Uses DS Library} or \textit{Does not use DS Library} (the three categories are mutually exclusive).

\begin{figure}[tpb]
  \centering
  \includegraphics[width=0.8\linewidth]{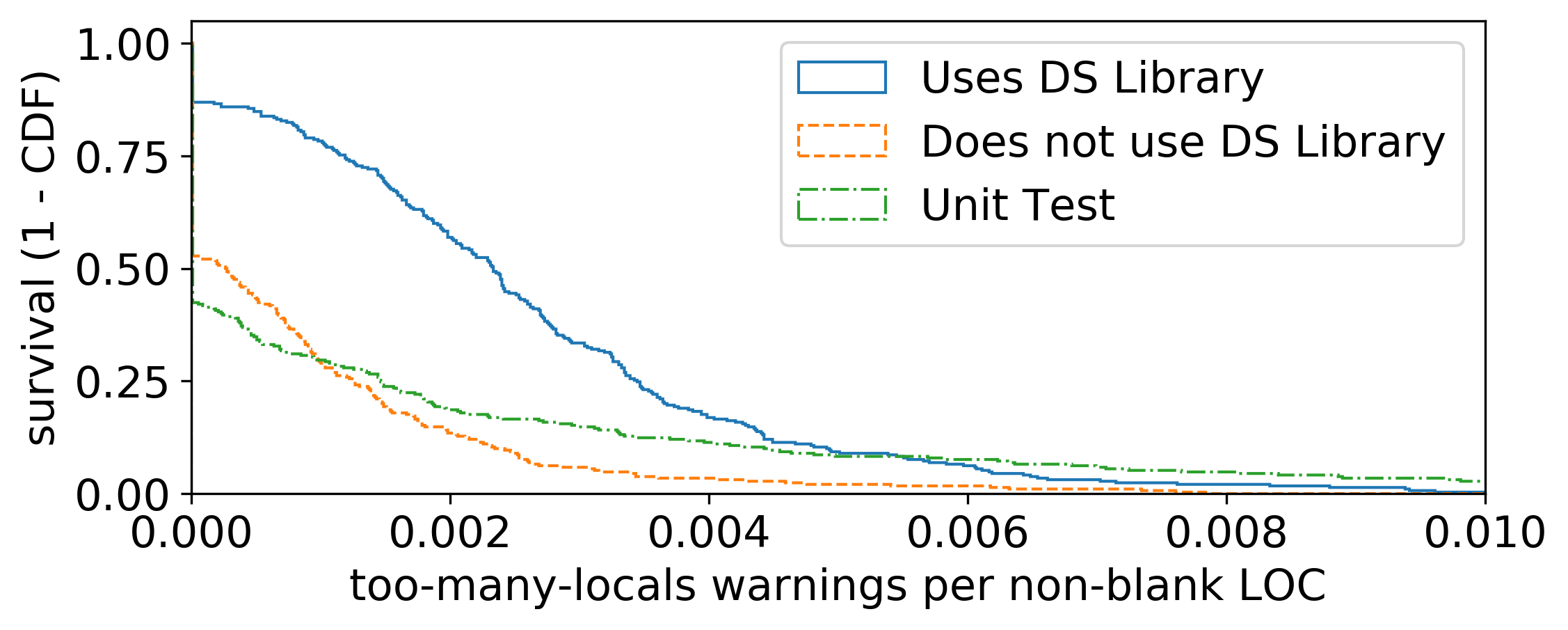}
  \includegraphics[width=0.8\linewidth]{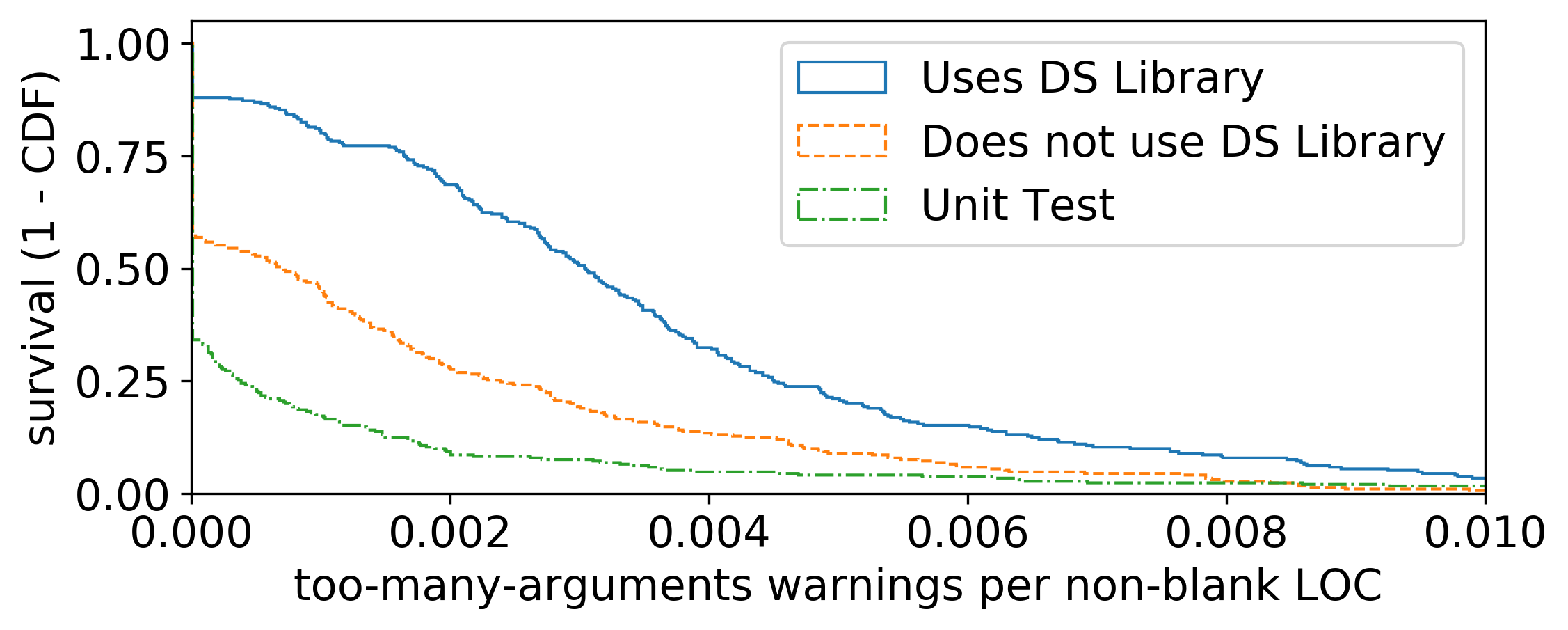}
  \includegraphics[width=0.8\linewidth]{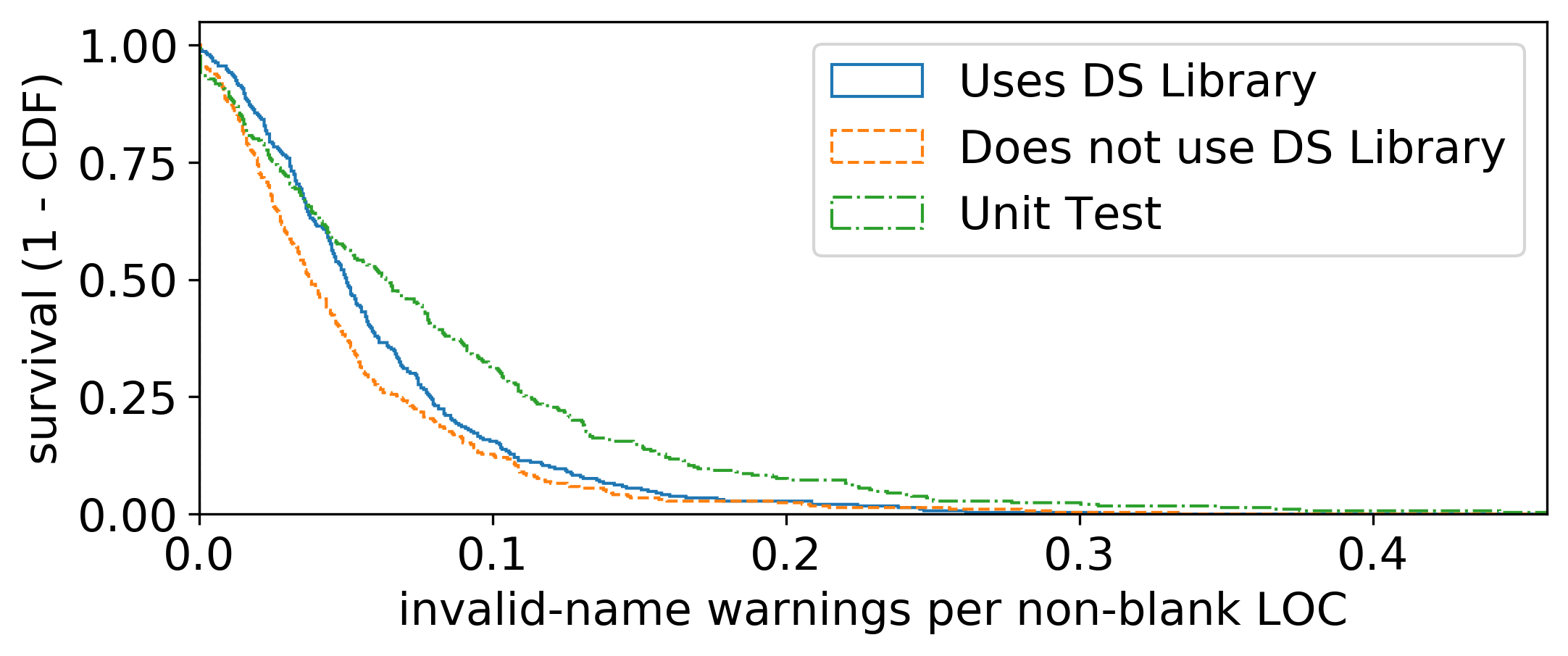}
  \includegraphics[width=0.8\linewidth]{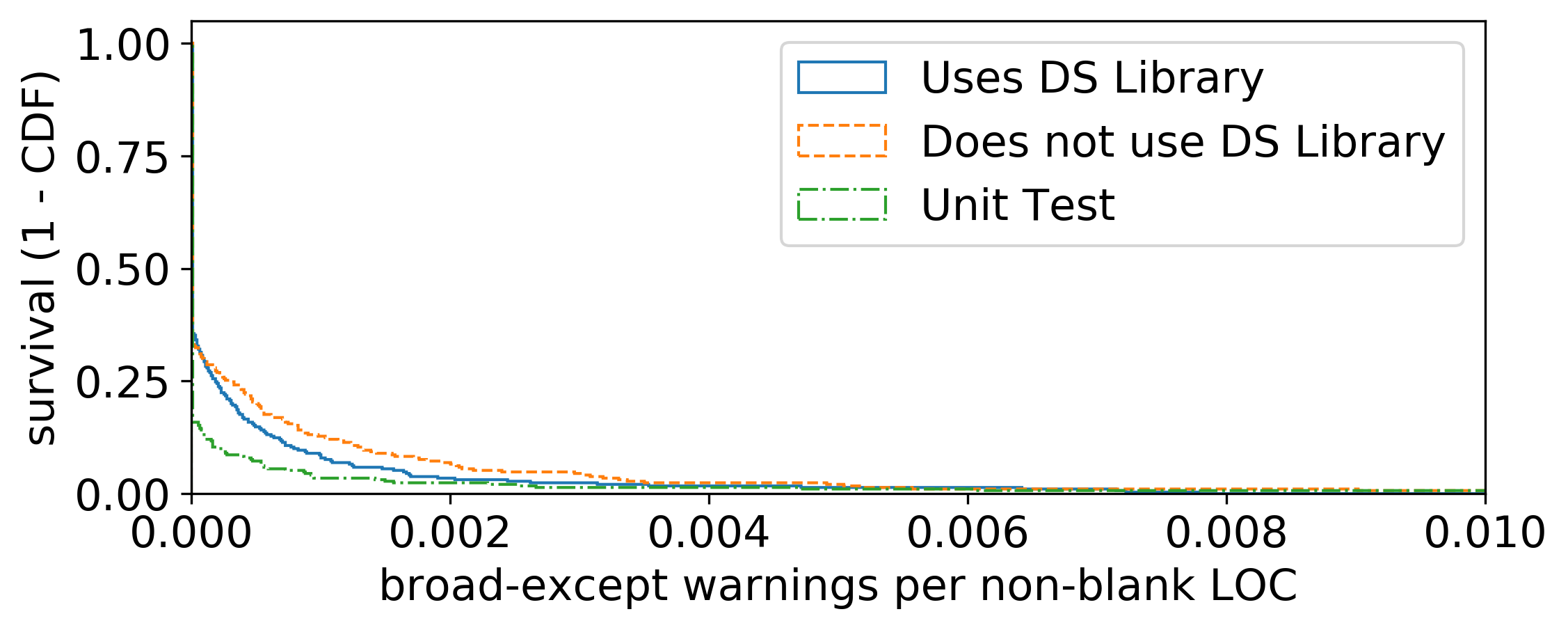}
  \caption{Survival plots for selection of coding standard violations within projects, categorised based on module imports.}
  \label{fig:freq-pylint-modules}
\end{figure}

To prevent the analysis from becoming swamped by large repositories with many modules, an intermediate average is calculated per category (\textit{Unit Test}, etc.) for each repository, such that the repository contributes one data point per category to the final distribution. For this research question, we only analyse Data Science repositories that contain a combination of all three categories (290 repositories). The final breakdown of the results is presented in \autoref{fig:freq-pylint-modules}.

The plot of \textit{too-many-locals} confirms that the violations are indeed coming from modules that import a Data Science library. However, other code in the repository does not appear to be affected.

The plot of \textit{invalid-name} shows that the naming violations are frequent in modules that imports Data Science libraries. However, other code in the repository, such as unit tests, also seem to be affected. This indicates that the naming violations spread throughout Data Science repositories.

The majority of code was defined in modules that imported Data Science libraries (median of 3706 non-blank LOC) in contrast to unit testing testing code (median 746 non-blank LOC) and other code (median 1604 non-blank LOC).

\begin{framed}
\textbf{RQ3 summary.} Overuse of local variables appears to be confined to modules that directly import Data Science libraries. In contrast, non-conforming variable names in Data Science projects occur throughout the entire project. The majority of Data Science code is defined in modules that import Data Science libraries.
\end{framed}




\section{Discussion}
\label{sec:dicussion}

In this section we use the technical domain as a lens to interpret our findings \cite{Barnett2017} as our study focused on the specific technical domain of Data Science. 
Our study found that Data Science projects are forked more often and have lower number of contributors which would indicate greater software reuse amongst Data Scientists. This is consistent with Data Science specific tasks such as data preparation, statistical modelling, and training machine learning models.  

Compared to other projects of similar community acceptance and maturity, Data Science projects showed significantly higher rates of functions that use an excessive number of parameters (\textit{too-many-arguments}) and local variables (\textit{too-many-locals}). 

Our module-level analysis confirms that this appears in code that imports machine learning libraries, but does not appear to affect other modules in the application. Despite the excessive number of local variables and parameters, the average cyclomatic complexity of functions in Data Science projects appeared similar to that of non-Data Science projects. This suggests that machine learning frameworks and libraries are effective in helping Data Scientists avoid branching and looping constructs, but not in managing the excessive number of variables associated with model parameters and configuration.

Variable naming conventions in Data Science projects also differ from those used in other projects, as evidenced by the higher frequency of Pylint  (\textit{invalid-name}) warnings. In contrast to the other warnings, the variable naming differences appear to permeate the entire Data Science code base.

\subsection{Implications}
In this subsection we present the implications of our research. 

\subsubsection{Software teams should be aware that Data Scientists may be following unconventional coding standards.} Our findings indicate that Data Science code follows different coding standards so this information should be taken into consideration when establishing team wide standards. 

\subsubsection{Further investigation to evaluate the case for Data Science aware coding standards needed.} Answers to RQ2 and RQ3 indicate that Data Science projects do not follow coding conventions to the same extent as a non-Data Science project. This raises at least two additional research questions `Do Data Scientists follow Data Science specific coding standards?' and `Why does conformance to coding standards differ between Data Science projects and non-Data Science projects?' To approach the first question we plan to do a qualitative analysis of the data collected in this paper to derive potential Data Science specific coding standards before an evaluation with practitioners. To better understand Data Scientists attitude to coding standards we propose a large scale survey.

\subsubsection{Study the impact of variable naming conventions in Data Science projects on readability.} A recent study indicated that short variable names take longer to comprehend \cite{hofmeister2019shorter} which may not hold for Data Science code when trying to comprehend the details of an algorithm implementation written to resemble math notation. Our study found that Data Science projects follow different coding standards than non-Data Science projects (RQ2) which potentially impacts readability. 

\subsubsection{Develop richer tools to support empirical software engineering research involving Python source code.} We found that Radon hangs on large Python files, and encountered multiple issues with Pylint as described in \autoref{sec:threats}. With Python's dominance in the Data Science space and continuing popularity in the Open-Source community \cite{Bafatakis2019} the need for tools to aid empirical software engineering research is going to grow. 

\subsubsection{Further study of conformance to Python idioms by Data Scientists.} The Python community is well known for the `Zen of Python' a set of high-level coding conventions and recent work has set about cataloguing these conventions \cite{Alexandru2018}. While our work indicated a discrepancy between coding standards in Data Science projects compared to non-Data Science projects, Python specific idioms not detected by Pylint may still have frequently been used.

\section{Threats to Validity}
\label{sec:threats}

\subsection{Internal validity}

\subsubsection{Confounding variables}
As seen in \autoref{fig:distributions}, our dataset includes a diversity of projects with a wide spread of lines of code. As such, it is vital to correct for the number of lines of code in the project when reporting the number of code style violations. We do this by reporting the number of violations per line of code. However, this may not be suitable for all warning types. For example, as imports only occur once at the top of the file, import related errors may not increase linearly with the number of lines of code (in our analysis we exclude import errors). Similarly, for errors related to functions (e.g. ``too-many-arguments''), it may be more appropriate to normalise by the number of functions rather than the lines of code.

\subsubsection{Assumption of independence}
An assumption in our calculation of statistical significance is that the data-points are independent. This assumption can be violated if the data contains duplicated repositories. As a guard against this, we excluded forked repositories, as noted in \autoref{sec:non-data-sci-selection}. However, during our investigation, we came across some instances where the exact same code violation was repeated across multiple repositories due to partial sharing of code that had been copied from another repository then modified. Another threat is that the same developer may be involved in multiple projects. As such, the probability that the patterns observed in our paper are coincidental (i.e. false positives) may be higher than suggested by the p-values due to projects that share code and/or authors.

\subsubsection{Robustness to outliers}
Some repositories contained non-project related code, such as dependencies copied directly into the repository, or even entire Python virtual environments. Furthermore, some repositories encoded data as Python files, leading to source files with over 50,000 lines of code. Only the maximum disk limit enforced by GitHub limits what a repository can contain, so any single repository has the potential to arbitrarily distort averages though committing irrelevant directories containing large Python files. To limit the extent to which any one repository can affect the results, \autoref{sec:pylint-ml} treats all repositories with equivalent weight (regardless of how many files they contain), and reports the median as this is more robust to outliers than the mean (but unfortunately is not suited for rare error types, for which the median is zero). The module-level analysis in \autoref{sec:module-level-analysis} also limits the impact of large repositories containing irrelevant modules by averaging over all files in a repository of a certain category (\textit{Unit Test}, \textit{Uses DS Library}, or \textit{Does not use DS Library}) such that each repository only contributes a single data-point for each category. Unfortunately however, the module-level analysis may be affected by uneven group sizes (e.g. if the \textit{Uses DS Library} category contains longer/more files that the \textit{Does not use DS Library} category, then there is a greater chance that it will contain at least one line that produces a warning, and thus has a greater chance of producing a positive, albeit small, warnings per LOC score). As such the distributions presented in this section need to be interpreted with caution; the median of the resultant distribution may be distorted due to uneven group sizes, but the mean of the resultant distribution (mean of means) will not be biased by group size.


\subsubsection{Validity of Pylint}


Pylint uses the current Python environment in order to resolve imports and detect coding errors such as calling a non-existent method or passing the wrong number of arguments. As such, the Pylint results may be influenced by which Python packages were installed in the Python environment---for example, if the project depends on a newer version of a library than what is installed in the Python environment, it may lead to warnings related to use of newer methods that do not appear in the installed version of the library. However, given the diversity of projects analysed (some of which did not specify dependencies, or only did so via a README file), it was impractical to reliably determine and install dependencies for the projects analysed. To ensure a fair comparison, we used a clean Python environment containing only essential packages needed for data extraction (\textit{findimports}, \textit{pylint}, \textit{radon}, and their dependencies). A consequence of this decision is that certain Pylint checks (such as \textit{import-error}) should be ignored.


As Pylint is popular in the Python community, projects that use Pylint (or a similar linting library) as part of their Continuous Integration (CI) are likely to have few (or no) warnings. Furthermore, source files may include ``\# pylint:'' comments that can instruct Pylint to ignore warnings for that file.  Thus our results may reflect whether projects used a linting system rather than developer commitment to code quality. However, Pylint identified at least one warning in nearly all repositories, indicating that Pylint is either not being widely used by repositories in the corpus, or that projects do not have all Pylint checks turned on.

Pylint also has known bugs relating to false positives. The Pylint issue tracker currently contains 104 open bugs (and 318 closed bugs) reporting or relating to a ``false positive''. As this paper examines differences in the number of code standard violations, false positives will not undermine the analysis if they are triggered consistently across all projects. Unfortunately however, false positives can be triggered by Python features such as type hints, leading to counter intuitive results where use of features such as type hints (designed to promote better code) has the potential to trigger more warnings.




\subsubsection{Identification of Data Science repositories} We relied on \citeauthor{Biswas2019} \cite{Biswas2019} for the identification of Data Science repositories, and negated their query to identify non-Data Science repositories. However, as an automated approach, it is possible that some traditional projects appear in the Data Science corpus, and vice~versa. Nevertheless, as our study focuses on examining the differences between these groups, the observed differences are still meaningful so long as the Data Science corpus contains a higher proportion of Data Science projects than the non-Data Science corpus. The effect of imperfect identification will be to reduce the observed strength of the differences. In future work we intend to manually perform a qualitative analysis of the corpora to form a more complete picture of their contents.

\subsection{External validity}

Our analysis was limited to Open-Source Data Science projects on GitHub. Additional research is needed to test whether our results generalise to closed-source Data Science projects undertaken for commercial reasons, in which there is the ability for managers to enforce top-down processes across the entire development team.


\section{Conclusion}
\label{sec:conclusion}

In this study we conduct the first large-scale empirical investigation to examine the extent to which Open-Source Data Science projects follow coding standards. Our results provide empirical evidence in support of our hypothesis that Data Science projects show differences in coding standards when compared to other projects. Specifically, we found that Open-Source Data Science projects in the corpus contain twice as many cases of functions/methods with excessive numbers of local variables and also contain more frequent violations of variable naming standards.

Through pursuing this line of research, we aim to identify barriers to communication within inter-disciplinary teams arising from differing norms, and restore the sense of shared ownership over the entire end-to-end Data Science workflow. Future work will involve qualitative analysis to identify causal links for our findings and an evaluation with practitioners to determine practically relevant coding standard violations. Our research provides an initial indication that the technical domain of Data Science may require specific coding standards. These coding standards could be used to inform development of a software linting tool tailored to the needs of Data Science projects.

\begin{acks}
The authors would like to thank Nikhil Bhat and the Surround development team for assistance with extracting the project metrics used in this paper, and acknowledge the Pylint project contributors. 
\end{acks}

\bibliographystyle{ACM-Reference-Format}

\bibliography{references,references2}

\end{document}